\documentstyle[12pt,epsfig,amsmath]{article}

\begin{document}

\title{Role of the $N^*(1440)$ resonance in the $pp \to pn  \pi^+$ reaction}
\author{Zhen Ouyang~$^{1,3,4)}$\thanks{Corresponding author. ouyangzh@impcas.ac.cn; Tel: +86-10-88236162;
fax: +86-10-88233085}, Ju-Jun
Xie~$^{2,3,4)}$\thanks{xiejujun@mail.ihep.ac.cn}, Bing-Song
Zou$^{2,3,4)}$\thanks{zoubs@mail.ihep.ac.cn}, Hu-Shan
Xu~$^{1,3,4)}$\thanks{hushan@impcas.ac.cn}  \\
1) Institute of Modern Physics, CAS, Lanzhou 730000, China  \\
2) Institute of High Energy Physics, CAS, Beijing 100049, China  \\
3) Graduate University of Chinese Academy of Sciences, Beijing 100049, China \\
4) Theoretical Physics Center for Science Facilities, CAS, Beijing
100049, China}
\date{}
\maketitle

\begin{abstract}
New measurement by CELSIUS-WASA Collaboration on the $pp \to pn
\pi^+$ reaction reveals clear evidence for the presence of the Roper
resonance $N^*(1440)$ which has been ignored in previous theoretical
calculations. In this article, based on an effective Lagrangian
approach and available knowledge on the Roper resonance, we
investigate the role of the Roper resonance for the $pp\to p n
\pi^+$ reaction. It is found that the contribution from the Roper
resonance $N^*(1440)$ becomes significant for kinetic energy above
1.1 GeV, consistent with the new experimental observation. The
t-channel $\sigma$-meson exchange is dominant for the production of
the Roper resonance.
\end{abstract}

{\it keywords}: $pp$ collision; Pion production; $N^*(1440)$
resonance; Effective Lagrangian approach

\section{Introduction}
Ever since the first even-parity excited state of the nucleon, the
Roper resonance $N^*(1440)$ was first deduced by $\pi N$ phase shift
analysis, its structure has been arousing people's interests
intensely all the time---it is lighter than the first odd-parity
nucleon excitation, the $S_{11}(1535)$, and also has a significant
branching ratio into two pions. Up to now, although the existence of
the Roper resonance is well established (4-star ranking in the
particle data book), its properties, such as mass, width and decay
branching ratios etc., still suffer large experiment
uncertainties~\cite{pdg2006}. In classical quark models, the Roper
resonance has been associated with the first spin-parity $J^P=1/2^+$
radial excited state of the nucleon~\cite{qmodel1,qmodel2,qmodel3,
qmodel4}. In the bag model~\cite{meissnernpa430} and in the Skyrme
model~\cite{hajdukplb140}, a surface oscillation, also called
breathing mode, has been predicted to interpret the Roper resonance
as a monopole excitation of the nucleon. Furthermore, it has also
been supposed to relate to a hybrid nature which means a gluonic
excitation state of the nucleon~\cite{hybrid1,hybrid2,hybrid3}. The
recent theoretical works~\cite{dymodel1,dymodel2} found that the
Roper resonance was dynamically generated from the meson-nucleon
interactions. Nevertheless, the model predictions always reach
either larger value for the Roper resonance mass or much smaller one
for its width and also meet difficulties to explain its
electromagnetic coupling~\cite{saran}.

In the early years, our knowledge on Roper resonance was mainly
based on $\pi N$ and $\gamma N$ experiments, however, it is excited
so weakly thus buried by the strong $\Delta$ peak in these
experiments. Another difficulty in extracting the Roper resonance
information from these experiments is the isospin decomposition of
$1/2$ and $3/2$~\cite{workman}. In Refs.~\cite{zounpa675,zouepja11},
it was pointed out that the decays of $J/\psi \to \bar{N}N\pi$ and
$J/\psi \to \bar{N}N\pi\pi$ provide an ideal place for studying the
properties of nucleon resonances, since in these processes the $\pi
N$ and $\pi\pi N$ systems are limited to be pure
isospin-$\frac{1}{2}$ due to isospin conservation. The result from
BES Collaboration on $J/\psi \to p\pi^- \bar{n} +$ c.c. reaction
showed a clear $N^*$ peak in the $N\pi$ invariant mass spectrum
around 1360 MeV/c$^2$ which gave the first direct observation of the
Roper resonance peak in the $\pi N$ invariant mass
spectrum~\cite{besroper}. Nevertheless, since the Roper resonance
has a strong coupling to $\sigma N$, studying the Roper resonance
from other production processes, $i.e.$, $\sigma N \to N^*(1440)$,
is necessary, as also suggested in Refs.~\cite{hiren,ruso}.

Recently, the CELSIUS-WASA Collaboration performed a measurement on
the $pp\to p n \pi^+$ reaction with proton beam of kinetic energy
T$_\text P=1.3$ GeV. They also observed a clear resonance structure
in the invariant $n\pi^+$ mass spectrum for the Roper resonance at
$M \approx 1360$ MeV/c$^2$ with a width of about 150
MeV~\cite{celsiusroper,nstar2007}. These values agree well with the
BES result~\cite{besroper} and the pole position of the Roper
resonance from $\pi N$ phase shift analyses~\cite{pdg2006}. The
$pp\to p n \pi^+$ reaction has the largest cross section for $pp$
collision in the intermediate energy region around T$_\text P=1.3$
GeV. It can be further studied by WASA-at-COSY and HIRFL-CSR
facilities with much higher precision and statistics. This reaction
opens a new window for studying the Roper resonance.

However, up to now, theoretical study on this channel in the
intermediate energy region is scarce. In Ref.~\cite{engelnpa603},
Engel {\sl et al.} performed a fully covariant calculation for the
total cross section of the $pp \to p n\pi^+$ reaction. However,
unaware of the significant role of the Roper resonance for this
reaction, their calculations only included the $\Delta$ resonance
and the off-shell nucleon pole. Furthermore, in their calculations
they did not consider the final-state interaction of $pn$. In
Ref.~\cite{kundu}, the authors performed their calculations for
T$_\text P=0.8$ GeV by including only the $\Delta$ resonance.

Inspired by the new observation of the Roper resonance by the
CELSIUS-WASA Collaboration~\cite{celsiusroper,nstar2007},  here we
investigate the role of the Roper resonance for the $pp\to p n
\pi^+$ reaction, based on an effective Lagrangian approach and
available knowledge on the Roper resonance.

In next section, we will give the formalism and ingredients for our
calculation. Then numerical results and discussion are given in
Sect.3.

\section{Formalism and ingredients}

We study the $pp \to pn\pi^+$ reaction in an effective Lagrangian
approach. All the basic Feynman diagrams involved for the reaction
are depicted in Fig.~\ref{diagram}. The so-called ``pre-emission"
diagrams~\cite{engelnpa603} are not shown. But their contributions
are included in actual calculations. In addition to the
$\Delta(1232)$ and the off-shell nucleon pole included in previous
works~\cite{engelnpa603,kundu,hudoprc18,shyamplb426}, the Roper
resonance is added in our calculation. For the production of the
Roper resonance and the intermediate virtual nucleon, the $\pi^0$
and $\sigma$-meson exchanges are included while the $\rho$-meson
exchange is found to give negligible contribution and is dropped.
For the production of the $\Delta(1232)$ resonance, we only include
$\pi$-meson exchange since iso-scalar meson exchanges cannot
contribute and according to Ref.~\cite{kundu} any attempt to include
the $\rho$-meson exchange worsens the agreement with experiments of
$pp\to n\Delta^{++}$ reactions. A Lorentz covariant
orbital-spin(L-S) scheme \cite{zouprc03} are used for the effective
$\Delta\pi N$, $N^{*}\sigma N$, and $N^*\pi N $ vertices.

\subsection{Meson-Baryon-Baryon (Resonances) vertices}

The effective Lagrangian densities involved for describing the $\pi
NN$ and $\sigma NN$ vertices are,
\begin{equation}
{\cal L}_{\pi N N}  = - \frac{f_{\pi N N}}{m_{\pi}} \overline{u}_{N}
\gamma_5 \gamma_{\mu} \vec\tau \cdot \partial^{\mu} \vec\psi_{\pi}
u_N , \label{piNN}
\end{equation}
\begin{equation}
{\cal L}_{\sigma N N}  =  g_{\sigma N N}  \overline{u}_{N}
\psi_{\sigma} u_N. \label{sigNN}
\end{equation}

At each vertex a relevant off-shell form factor is used. In our
computation, we take the same form factors that used in the
well-known Bonn potential model~\cite{mach}:
\begin{equation}
F^{NN}_M(k^2_M)=\frac{\Lambda^2_M-m_M^2}{\Lambda^2_M- k_M^2},
\end{equation}
where $k_M$, $m_M$ and $\Lambda_M$ are the 4-momenta, mass and
cut-off parameter for the exchanged meson ($M$), respectively. In
the Bonn meson-exchange model~\cite{mach} for the nucleon-nucleon
interaction, $k^0_M=0$, hence $k_M^2=-|\vec k_M|^2$ is negative. For
the $pp\to\pi^+pn$, it can easily verify that for the t-channel
meson exchange, the $k_M^2$ still keeps negative. Therefore the
off-shell form factors here always give a suppression factor. The
coupling constants and the cut-off parameters are taken as the
following ones~\cite{mach,tsushima1,tsushima2,tsushima3,sibi1,xie}:
$g_{\pi NN}^2/4\pi = 14.4$, $\Lambda_{\pi}$ = 1.3 GeV, $g_{\sigma
NN}^2/4\pi = 5.69$ and $\Lambda_{\sigma}$ = 1.7 GeV. Note that the
constant $g_{\pi NN}$ is related to $f_{\pi NN}$ of Eq. (\ref{piNN})
as $g_{\pi NN}=(f_{\pi N N}/m_{\pi})2m_N$~\cite{scadron}.

\begin{figure}[htpb]
\begin{center}
\includegraphics[scale=0.6]{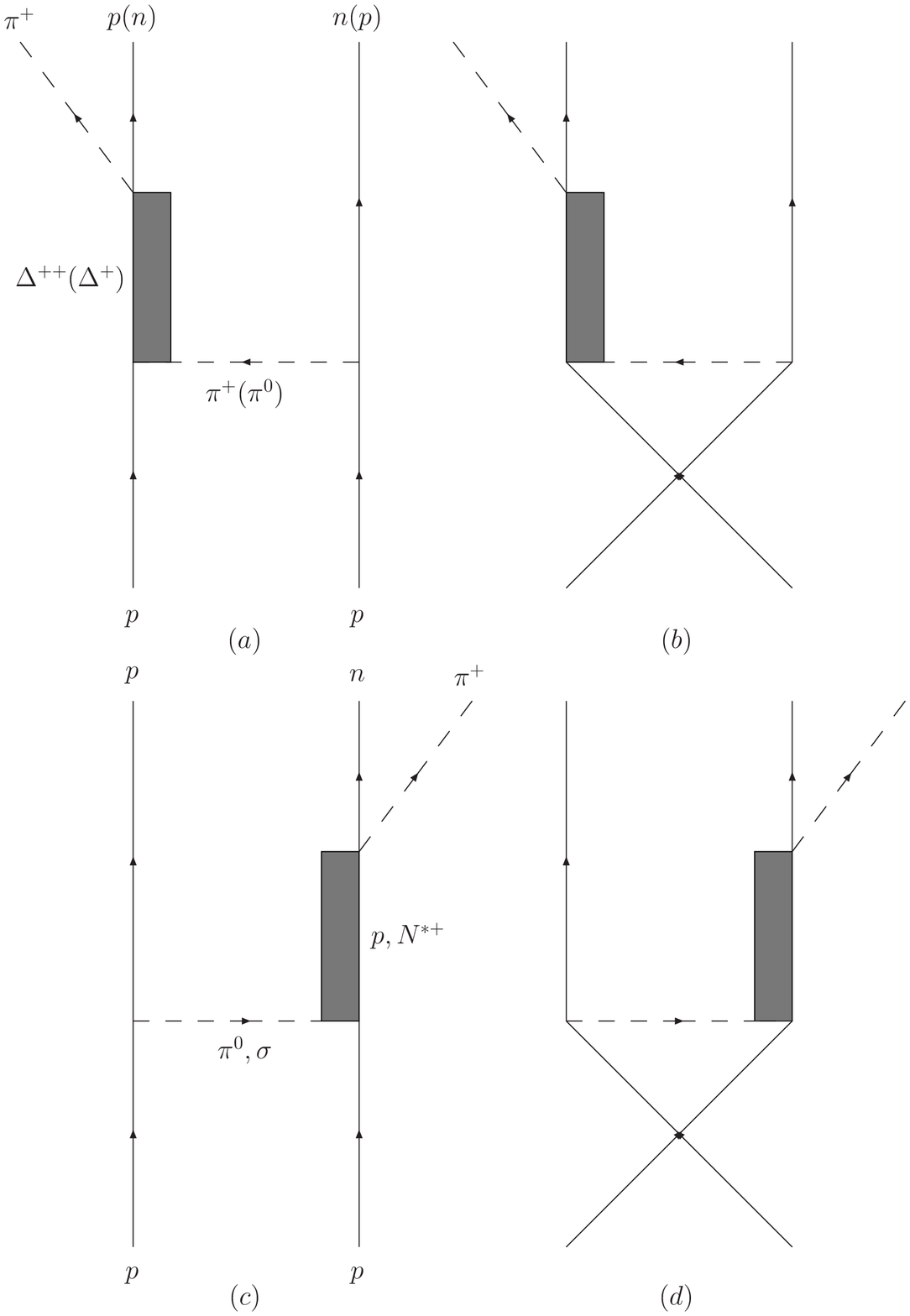} \vspace{0.3cm}
\caption{} \label{diagram}
\end{center}
\end{figure}

To calculate the amplitudes of diagrams in Fig.~\ref{diagram} with
the resonance model, we also need to know interaction vertices
involving $\Delta(1232)$ and $N^*(1440)$ resonances. In
Ref.~\cite{zouprc03}, a Lorentz covariant orbital-spin scheme for
$N^* NM$ couplings has been described in detail, which can be easily
extended to describe the $\Delta(1232) N \pi$, $N^*(1440)N\pi$ and
$N^*(1440)N\sigma$ couplings that appear in the Feynman diagrams
depicted in Fig.~\ref{diagram}. By using that scheme, we can easily
obtain the effective $\Delta(1232)N \pi$,$N^*(1440)N \pi$, and
$N^*(1440)N \sigma$ couplings:
\begin{eqnarray}
{\cal L}_{\pi N \Delta(1232)} &=& g_{\Delta(1232) N \pi}
\overline{u}_{N}  \partial^{\mu} \psi_{\pi} u_{\Delta(1232)\mu} + \text{h.c.}, \label{1232pi} \\
{\cal L}_{\pi N N^*(1440)} &=& g_{N^*(1440)N \pi}
\overline{u}_{N} \gamma_5 \gamma_{\mu}  \partial^{\mu} \psi_{\pi} u_{N^*(1440)} + \text{h.c.}, \label{1440pi} \\
{\cal L}_{\sigma N N^*(1440)} &=& g_{N^*(1440) N \sigma}
\overline{u}_{N} \psi_{\sigma} u_{N^*(1440)} + \text{h.c.},
\label{1440sig}
\end{eqnarray}
with $u_N$, $u_{N^*(1440)}$, $u_{\Delta^*\mu}$, $\psi_{\pi}$, and
$\psi_{\sigma}$ as the  wave functions for the nucleon, $N^*(1440)$
resonance, $\Delta^*(1232)$ resonance, $\pi$ and $\sigma$-meson,
respectively.

For the relevant vertices involving $\Delta(1232)$ and $N^*(1440)$
resonances, the off-shell form factors are adopted as below:
\begin{equation}
F_M(k^2_M)=(\frac{\Lambda^{*2}_M-m_M^2}{\Lambda^{*2}_M- k_M^2})^n,
\end{equation}
where n=1 for the Roper resonance and n=2 for the $\Delta(1232)$
resonance ~\cite{hiren,engelnpa603,hern}, with $\Lambda^*_M$ being
the corresponding cut-off parameters which are:
$\Lambda^{\Delta}_{\pi} = 0.59$ GeV, $\Lambda^{N^*(1440)}_{\pi} =
1.3$ GeV~\cite{hiren} and $\Lambda^{N^*(1440)}_{\sigma} = 1.5$
GeV~\cite{hern}. Here a quite severe cutoff $\Lambda^{\Delta}_{\pi}$
is needed to reproduce data, in consistent with the previous
study~\cite{kundu} of $pp\to n\Delta^{++}$ reactions.

It is worth noting that we also take the virtuality of the
intermediate nucleon into account by introducing a form factor which
is taken as in Refs.~\cite{pennerprc66,shkprc72,feusterprc58}
\begin{equation}
F(q)=\frac{\Lambda^4_N}{\Lambda^4_N+ (q^2-m_N^2)^2},
\end{equation}
with $\Lambda_{N}$= 0.6 GeV.

\subsection{Coupling constants for the intermediate resonances}

The $\Delta(1232)N\pi$ and $N^*(1440)N\pi$ coupling constants are
determined from the experimentally observed partial decay widths of
the $\Delta$ and Roper resonances. The general formula for the
partial decay width of $\Delta(1232)$ or $N^*(1440)$ resonance
decaying into a nucleon and a pion is as the following
\begin{equation}
d \Gamma =\overline{|{\cal M}_{R\to N\pi} |^2} (2\pi)^4
\delta^4(q_R-p_N-p_\pi) \frac{d^3 p_N}{(2\pi)^3} \frac{m_N}{E_N}
\frac{d^3p_\pi}{(2\pi)^3} \frac{1}{2E_\pi},
\end{equation}
where ${\cal M}_{R\to N\pi }$ stands for the total amplitude of the
initial resonance decaying to a nucleon and a pion, the $q_R$, $p_N$
and $p_\pi$ are the 4-momenta of the three particles, $E_N$ and
$E_\pi$ are the corresponding energies. With the effective
Lagrangians described by Eq. (\ref{1232pi}) and Eq. (\ref{1440pi}),
the partial decay widths can be calculated
\begin{equation}
\Gamma_{N^*(1440)\to N \pi}=\frac{3g^2_{N^*(1440) N
\pi}p_N^{cm}}{4\pi
}[\frac{m^2_\pi(E_N-m_N)}{M_{N^*(1440)}}+2(p_N^{cm})^2]
\label{1440d}
\end{equation}
with
\begin{equation}
p_N^{cm}=\sqrt{\frac{(M^2_{N^*(1440)}-(m_N+m_{\pi})^2)
(M^2_{N^*(1440)}-(m_N-m_{\pi})^2)}{4M^2_{N^*(1440)}}},
\end{equation}
\begin{equation}
E_N=\sqrt{(p_N^{cm})^2+m^2_N},
\end{equation}

\begin{equation}
\Gamma_{\Delta^*(1232) N \pi}=\frac{g^2_{\Delta^*(1232) N
\pi}(m_N+E_N)(p_N^{cm})^3}{12 \pi M_{\Delta^*(1232)}}, \label{1232d}
\end{equation}
with $p_N^{cm}$ here the momentum of the nucleon in the rest frame
of $\Delta(1232)$.

For the $N^*(1440) N \sigma$ coupling constant, we get it from the
partial decay width of $N^*(1440) \to N \sigma \to N \pi \pi$, which
is given by
\begin{equation}
d \Gamma_{N^*(1440)  \to N \sigma \to N \pi \pi} = \overline{|{\cal
M}_{N^* \to N \sigma \to N \pi \pi}|^2} \frac{d^3p_1}{(2 \pi)^3}
\frac{m_1}{E_1} \frac{d^3p_2}{(2 \pi)^3} \frac{1}{2 E_2}
\frac{d^3p_3}{(2 \pi)^3} \frac{1}{2 E_3} (2 \pi)^4
\delta^4(p_{N^*}-p_1-p_2-p_3)
\end{equation}
with
\begin{equation}
{\cal M}_{N^* \to N \sigma \to N \pi \pi}  =  2g_{\pi \pi \sigma}
g_{N^* N \sigma} F^{N^* N}_{\sigma}(k^2_{\sigma}) \bar{u}_N(p_1,s_1)
\frac{i}{k^2_{\sigma}-m^2_{\sigma}+im_{\sigma}\Gamma_{\sigma}} p_2
\cdot p_3
 \bar{u}_{N^*}(p_{N^*},s_{N^*}),
\end{equation}
where ${\cal M}_{N^* \to N \sigma \to N \pi \pi}$ represents the
total amplitude of $N^*(1440)  \to N \sigma \to N \pi \pi$.
$p_{N^*}$ and $k_{\sigma}$ denote the 4-momenta of the $N^*(1440)$
resonance and the intermediate $\sigma$ meson; $m_{\sigma}$ and
$\Gamma_{\sigma}$ are the mass and full width of the $\sigma$ meson,
which we take $m_{\sigma}=550$ MeV and $\Gamma_{\sigma}=500$ MeV;
$p_1$, $m_1$, and $E_1$ stand for the 4-momenta, mass, and energy of
the nucleon; $s_1$ and $s_{N^*}$ the spin projection of the nucleon
and the $N^*(1440)$ resonance; $p_2$, $p_3$, and $E_2$, $E_3$ stand
for the 4-momenta and energy of the final two pions, respectively.
To calculate the amplitude of $\sigma \to \pi \pi$ appearing in the
total amplitude calculation, we use the following type of vertex
\begin{equation}
{\cal L}_{\pi \pi \sigma}  =  g_{\pi \pi \sigma}
\partial_{\mu}\vec\psi_{\pi} \cdot \partial^{\mu}\vec\psi_{\pi} \psi_{\sigma}, \label{sp}
\end{equation}
where the coupling constant $g_{\pi \pi \sigma}$ can be determined
from the partial decay width $\Gamma_{\sigma\to \pi\pi}$
\begin{equation}
d \Gamma_{\sigma \to \pi\pi} = \frac{1}{2m_{\sigma}}\overline{|{\cal
M}_{\sigma\to \pi\pi} |^2}  \frac{d^3p_1}{(2 \pi)^3} \frac{1}{2 E_1}
\frac{d^3p_2}{(2 \pi)^3} \frac{1}{2 E_2} (2 \pi)^4
\delta^4(p_{\sigma}-p_1-p_2),
\end{equation}
where ${\cal M}_{\sigma \to \pi \pi}$ represents the total amplitude
of $\sigma \to \pi \pi$. $p_1$, $p_2$, and $E_1$, $E_2$ stand for
the 4-momenta and energy of the final two pions, respectively.

With the experimental branching ratios \cite{pdg2006} and the above
formulae, we obtain all the corresponding coupling constants as
summarized in Table~\ref{table}.

\subsection{Propagators}

In our amplitude calculation of Feynman diagrams in
Fig.~\ref{diagram}, we also need propagators for the $\pi$ and
$\sigma$-meson, and the intermediate $\Delta(1232)$ and $N^*(1440)$
resonances with half-integer spin as well. For the $\pi$ and
$\sigma$-meson, the propagators are
\begin{equation}
G_{\pi/\sigma}(k_{\pi/\sigma})=\frac{i}{k_{\pi/\sigma}^2-m^2_{\pi/\sigma}},
\end{equation}
where $k_{\pi/\sigma}$ are the 4-momenta of $\pi$ and
$\sigma$-meson, respectively; $m_{\pi/\sigma}$ are their
corresponding masses.

The propagators for the $\Delta(1232)$ and $N^*(1440)$ resonances
can be constructed with their projection operators and the
corresponding Breit-Wigner factor~\cite{liang02}. The
spin-(n+$\frac{1}{2}$) resonance propagator can be written as
\begin{equation}
G^{n+\frac{1}{2}}(q)=P^{(n+\frac{1}{2})}\frac{2M_R}{q^2-M^2_R+iM_R\Gamma_R},
\end{equation}
where $1/(q^2-M^2_R+iM_R\Gamma_R)$ is the standard Breit-Wigner
factor; $M_R$, $q$ and $\Gamma_R$ are the mass, four momentum and
full width of the resonance, respectively. $P^{(n+\frac{1}{2})}$ is
the projection operator
\begin{equation}
P^\frac{1}{2}(q)=\frac { \not\!q  +M_R}{2M_R},
\end{equation}
\begin{equation}
P_{\mu\nu}^\frac{3}{2}(q)=\frac{(\not\! q + M_R )}{2M_R}\left[
g_{\mu \nu} - \frac{1}{3} \gamma_\mu \gamma_\nu - \frac{1}{3 M_R}(
\gamma_\mu q_\nu - \gamma_\nu q_\mu) - \frac{2}{3 M^2_R} q_\mu q_\nu
\right].
\end{equation}

With the projection operators, the propagators for the intermediate
$\Delta(1232)(\frac{3}{2}^+)$ and $N^*(1440)(\frac{1}{2}^+)$
resonances can be easily obtained as the following
\begin{equation}
G^{N^*(1440)}(q) =  \frac{\not\!q + M_{N^*(1440)}}{q^2 -
M^2_{N^*(1440)} + iM_{N^*(1440)} \Gamma_{N^*(1440)}}\,,
\label{spin1/2}
\end{equation}
\begin{equation}
G^{\Delta(1232)}_{\mu \nu}(q) =  \frac{P_{\mu \nu}(q)}{q^2 -
M^2_{\Delta(1232)} + iM_{\Delta(1232)} \Gamma_{\Delta(1232)}}
\label{spin3/2}
\end{equation}
with
\begin{eqnarray}
P_{\mu \nu}(q) & = &(\not\! q + M_{\Delta(1232)})  [ g_{\mu \nu} -
\frac{1}{3} \gamma_\mu \gamma_\nu - \frac{ \gamma_\mu q_\nu -
\gamma_\nu q_\mu}{3 M_{\Delta(1232)}} - \frac{2}{3
M^2_{\Delta(1232)}} q_\mu q_\nu].\label{pmunu}
\end{eqnarray}

For simplicity we use constant width in the Breit-Wigner (BW)
formula. The mass and width appearing in such constant-width BW
formula are very close to their corresponding pole positions. In
some detailed data fitting works, various energy dependent widthes
are used in the BW formula. The mass and width appearing in such BW
formula could then be very different from their corresponding pole
positions. Usually the pole positions are better determined than the
model dependent BW mass and width. For example, the BW mass and
width given by PDG~\cite{pdg2006} with energy dependent width are
$(1420\sim 1470)$ MeV and $(200\sim 450)$ MeV, respectively; while
the corresponding pole positions listed in PDG~\cite{pdg2006} are
$(1350\sim 1380)$ MeV and $(160\sim 220)$ MeV, respectively. Hence,
for parameters appearing in the constant-width BW propagators, we
adopt $M_{\Delta(1232)}=1.21$ GeV, $\Gamma_{\Delta(1232)}= 100$ MeV,
$M_{N^*(1440)}=1.365$ GeV and $\Gamma_{N^*(1440)}=190$ MeV, as given
by PDG for their corresponding pole positions~\cite{pdg2006}.

As for the intermediate virtual nucleon, we use
\begin{equation}
G^N(q)=\frac{\not\!q + m_N}{q^2-m^2_N}
\end{equation}
as its propagator, with $M_N$ and $q$ being the mass and 4-momenta
of the virtual nucleon, respectively.

\subsection{Amplitude and total cross section for $pp \to pn  \pi^+$ reaction}

In order to show the structure of total amplitude in terms of the
meson exchange and the intermediate resonances, we write down
explicitly the total amplitude in our calculation for the $pp \to pn
\pi^+$ reaction as the sum of sub-amplitudes
\begin{eqnarray}
{\cal M}& = & {\cal M}(\Delta(1232),\pi^+)+{\cal
M}(\Delta(1232),\pi^0)+{\cal M}(N^{*}(1440),\pi^0)+{\cal
M}(N^{*}(1440),\sigma)+ \nonumber\\
&& {\cal M}(p(938),\pi^0)+{\cal M}(p(938),\sigma), \label{ampl}
\end{eqnarray}
where on the right-hand side of the equation, the resonances and
mesons exchanged in the intermediate states are written inside the
bracket explicitly. Each sub-amplitude can be obtained
straightforwardly with effective couplings and propagators given in
former sections by following the Feynman rules. The explicit
expressions of the amplitudes are given in Appendix.

For the final-state-interaction(FSI) between the three outgoing
particles, we only need to consider $p$-$n$ FSI enhancement factor
near threshold, since the interaction between $n\pi^+$ and $p\pi^+$
are dominated by the s-channel intermediate resonances. To study
possible influence from the $p$-$n$ final state interaction, we
include it in our calculation by factorizing the reaction amplitude
as
\begin{eqnarray}
{\cal A}={\cal M}(pp \to p n \pi^+) T_{pn}, \label{fsiamp}
\end{eqnarray}
where ${\cal M}(pp \to p n \pi^+)$ is the primary production
amplitude as discussed above, $T_{pn}$ describes the $p$-$n$ final
state interaction, which goes to unity in the limit of no FSI. The
enhancement factor $T_{pn}$ is taken into account by means of the
general framework based on the Jost function
formalism~\cite{jostfsi1,jostfsi2} with
\begin{eqnarray}
T_{pn}=\frac{k+ i \beta}{k- i \alpha}. \label{fsi}
\end{eqnarray}
where $k$ is the internal momentum of $p$-$n$ subsystem, and the
$\alpha$ and $\beta$ are associated with the effective-range
parameters via~\cite{hinter}
\begin{eqnarray}
\alpha=\frac{1}{r}(1-\sqrt{1-\frac{2r}{a}}), ~~~~~
\beta=\frac{1}{r}(1+\sqrt{1-\frac{2r}{a}}).
\end{eqnarray}

According to the FSI analysis of Ref.~\cite{hinter} on the reaction
$pp \to p n\pi^+$ at the beam energy of $T_p=0.95$ GeV, the triplet
$^3S_1$ $np$ state dominates the $np$ invariant mass spectrum at the
low mass end with negligible singlet $^1S_0$ state contribution. So
in the present work, we assume the total amplitude with triplet
$^3S_1$ $np$ state dominating the $np$ near threshold region, and
multiply the total amplitude with the FSI factor of triplet $^3S_1$
$np$ state with the parameters adopted from Ref.~\cite{hinter}, {\sl
i.e.},
\begin{eqnarray}
a_t = 5.424 ~~\text{fm}, ~~~~~ r_t = 1.759 ~~\text{fm}.
\end{eqnarray}

Then the calculation of the invariant amplitude square $|\cal
A|^{\text {2}}$ and the cross section $\sigma (pp \to pn  \pi^+)$
are straightforward,
\begin{eqnarray}
d\sigma (pp\to p n \pi^+)=\frac{1}{4}\frac{m^2_p}{F} \sum_{s_i}
\sum_{s_f} |{\cal A}|^2\frac{m_p d^{3} p_{3}}{E_{3}} \frac{d^{3}
p_{\pi}}{2 E_{\pi}} \frac{m_{n} d^{3} p_{n}}{E_{n}} \delta^4
(p_{1}+p_{2}-p_{3}-p_{\pi}-p_{n}),  \label{eqcs}
\end{eqnarray}
with the flux factor
\begin{eqnarray}
F=(2 \pi)^5\sqrt{(p_1\cdot p_2)^2-m^4_p}. \label{eqff}
\end{eqnarray}
The factors 1/4 and $\sum_{s_i} \sum_{s_f}$ emerge for the simple
reason that the polarization of initial and final particles is not
considered. Since the relative phases among various meson exchanges
in the amplitude of Eq.~(\ref{ampl}) are unknown, the interference
terms are ignored in our concrete calculations.

In the amplitude given above, we have neglected the $pp$
initial-state-interaction (ISI) factor which has been used by many
authors in investigating $pp\to pp\eta$ reaction \cite{Naka,Wilkin}.
The role of the ISI in the production of a heavy meson from an NN
collision is basically to reduce the cross section near threshold by
an overall factor with little energy dependence, although with some
controversy about its absolute value. While the ISI factor with
$pp\to pp\eta$ reaction near threshold is relatively simple with
$^3P_0$ $pp$ initial state dominant, it becomes technically more
complicated for $pp\to NN\pi$ with several important initial partial
waves even at energies very close to the threshold. In practice, for
the pion production reactions, the ISI dumping effect is not
included explicitly in previous
calculations~\cite{engelnpa603,kundu} and is done by adjusting the
vertex form factors to reproduce the total cross sections. Since how
to treat this dumping factor does not influence our main conclusion
here, we follow the simple practical approach as these previous
calculations~\cite{engelnpa603,kundu} on the same reaction, although
the ISI could be taken into account by some more complicated
approaches, such as by including the partial wave ISI
factors~\cite{Naka,Wilkin} or distortions of wave
functions~\cite{Baru,horo,Koltun}.

\section{Numerical results and discussion}

\begin{figure}[htbp]
\begin{center}
\includegraphics[]%
{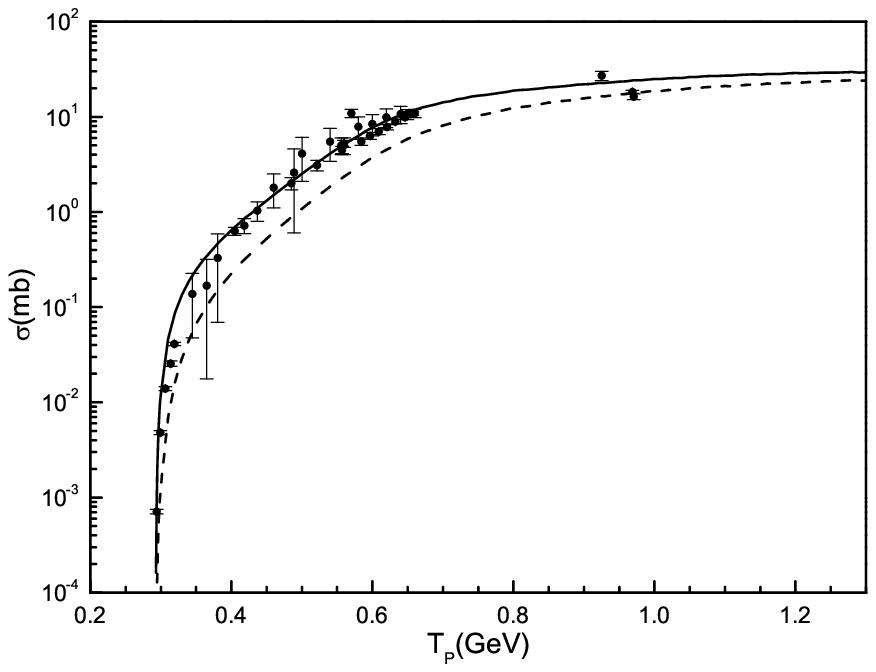}%
\caption{ } %
\label{tcs1}%
\end{center}
\end{figure}

\begin{figure}[htbp]
\begin{center}
\includegraphics[]{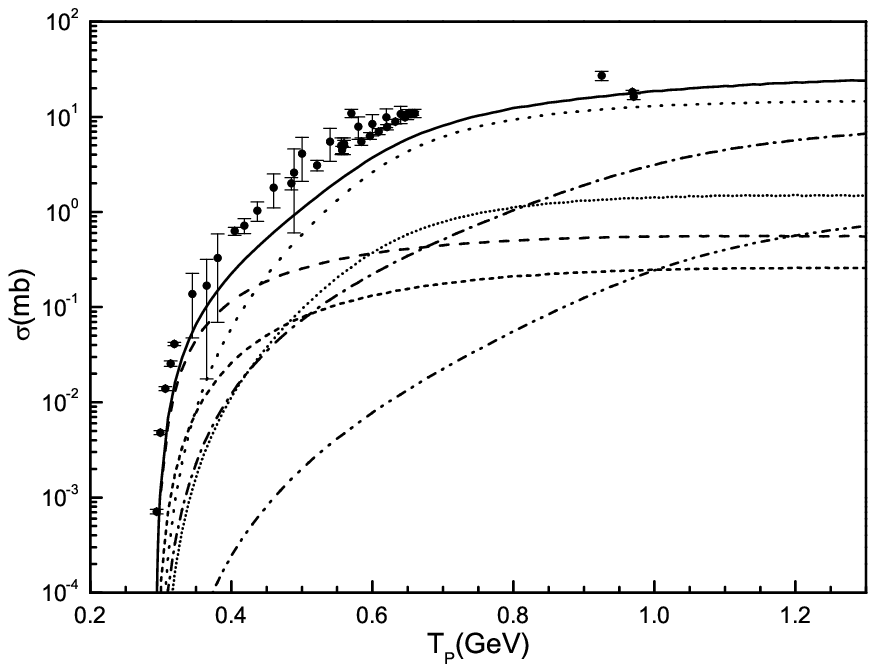}%
\caption{} \label{tcs2}%
\end{center}
\end{figure}

With the formalism and ingredients discussed in the former sections,
we evaluated the total cross section versus the kinetic energy of
the proton beam (T$_\text P$) for the $pp \to pn  \pi^+$ reaction by
using the code FOWL from the CERN program library, which is a
program for Monte Carlo multi-particle phase space integration
weighted by the amplitude squared. The results for T$_\text P$
ranging from 0.2 to 1.3 GeV are shown in Fig.~\ref{tcs1} and
Fig.~\ref{tcs2} along with experimental data~\cite{data1,data2} for
comparison.

In Fig.~\ref{tcs1}, the results obtained from with and without
including the $^3S_1$ $np$ FSI factor are represented by solid and
dashed curves, respectively. We can see our theoretical result
without FSI agrees well with the experimental data at beam energies
near 1.0 GeV. However, at lower beam energies,the calculated total
cross sections underestimate the data by a factor of 4 or more. In
view of the important role of FSI for the near-threshold
enhancement, we also tried to include it and found that the result
with FSI is indeed in excellent agreement with the experimental data
over a wide range of beam energies. Both the energy dependence and
the absolute magnitudes of the experimental data are very well
reproduced. Note that in reality, the $np$ $^3S_1$ partial wave
should dominate only at energies close to threshold. For energies of
a few hundred MeV above threshold, the contributions from higher
partial waves become important. The FSI effect due to the strong
$np$ $^3S_1$ interaction should become less important as energy
increases. Hence after taking this effect into account, one would
expect to get theoretical results closer to the solid curve at low
energies and to the dashed curve at higher energies in the
Fig.~\ref{tcs1}.

In Fig.~\ref{tcs2}, individual contributions corresponding to
$\Delta(1232)$ with $\pi^+$ and $\pi^0$ exchange, $N^*(1440)$ with
$\sigma$ and $\pi^0$ exchange, and nucleon pole with $\sigma$ and
$\pi^0$ exchange are shown in comparison with the experimental data
by dotted, short-dotted, dot-dashed, dot-dot-dashed, dashed, and
short-dashed curves, respectively. The contribution from the
$\Delta^{++}(1232)$ production by the $\pi^+$ exchange is found to
be dominant for T$_\text P$ between 0.6 and 1.1 GeV. The
$\Delta^{+}(1232)$ with $\pi^0$ exchange gives one order of
magnitude smaller contribution for the cross section than
$\Delta^{++}(1232)$ with $\pi^+$ exchange over the whole energy
region, simply due to smaller isospin factor of 1/9. In
Ref.~\cite{kundu}, by only including the contribution of the
$\Delta(1232)$ resonance the authors reproduced the experimental
points for present channel fairly well, however, visibly
underestimated the data points close to threshold. Actually, from
Fig.~\ref{tcs2}, one can see that the nucleon pole by $\sigma$ meson
exchange plays a more important role for beam energies close to
threshold. This is consistent with the previous calculation for the
$pp\to pp\pi^0$ reaction~\cite{horo}, where it is shown that the
prediction with only pion exchange~\cite{Koltun} underestimates the
data by a factor about 5. And there are more examples from other
reactions showing the significance of the sub-threshold nucleon pole
contribution, such as in $J/\psi\to \bar pn\pi^+$
\cite{besroper,liang2}. In contrast, contribution of the nucleon
pole by $\pi^0$ exchange is much smaller than by $\sigma$ exchange
over the whole energy region. Furthermore, it is worth noting that
contribution of the $N^*(1440)$ resonance by $\sigma$ exchange
becomes quite significant at beam energies above 1.1 GeV, larger
than all other contributions except $\Delta^{++}(1232)$
contribution, though it is negligibly small at lower beam energies.
The contribution of the $N^*(1440)$ by $\pi^0$ exchange is much
smaller by a factor of about 10 or more, which indicates that by no
means can it account for the clear Roper peak in the invariant mass
$M_{n\pi^+}$ spectrum. Anyway, the $\sigma$-meson exchange is found
to play a significant role, which has also been pointed out in a
previous study of $\alpha p$ reaction~\cite{hiren}. Possible
contribution from the neighborhood $N^*(1535)$ resonance is also
checked and is found to be negligible in the present energy region.

In our present calculation, we have not included the contribution
from $\pi N$ S-wave re-scattering process, which was found to be the
dominant production mechanism for the $\pi^+$ production at the
energies very close to the threshold~\cite{Lensky}. Since the main
interests of the present work is for the $N^*$ production at higher
energies and in our approach the uncertainties from FSI and ISI are
also quite large at the energies very close to the threshold, we
refer the readers who are interested in the very near threshold
regime to the more dedicated study of this regime by
Ref.~\cite{Lensky}.

In our concrete calculation, the interference terms between various
resonances are ignored because the relative phases from various
resonances are unknown. In order to get some feeling about the
importance of the interference terms, we calculated the magnitudes
of a few largest ones. At the high energy end $T_P=1.3$ GeV, the
largest two contributions come from $\Delta^{++}$ and $N^*(1440)$
resonances as shown in Fig.~\ref{tcs2}. For the cross section, the
contribution of the interference term from these two resonances is
found to be in the range of $\pm 5\%$ of the contribution from the
$\Delta^{++}$ resonance alone. At the lower energies around 0.4 GeV,
the largest two contributions from $\Delta^{++}$ and off-shell
nucleon are comparable, but the contribution from their interference
term is found to be only around $\pm 1\%$ to the total cross section
by integrating over the whole three-body phase space.

\begin{figure}[htbp]
\begin{center}
\includegraphics[scale=0.8]%
{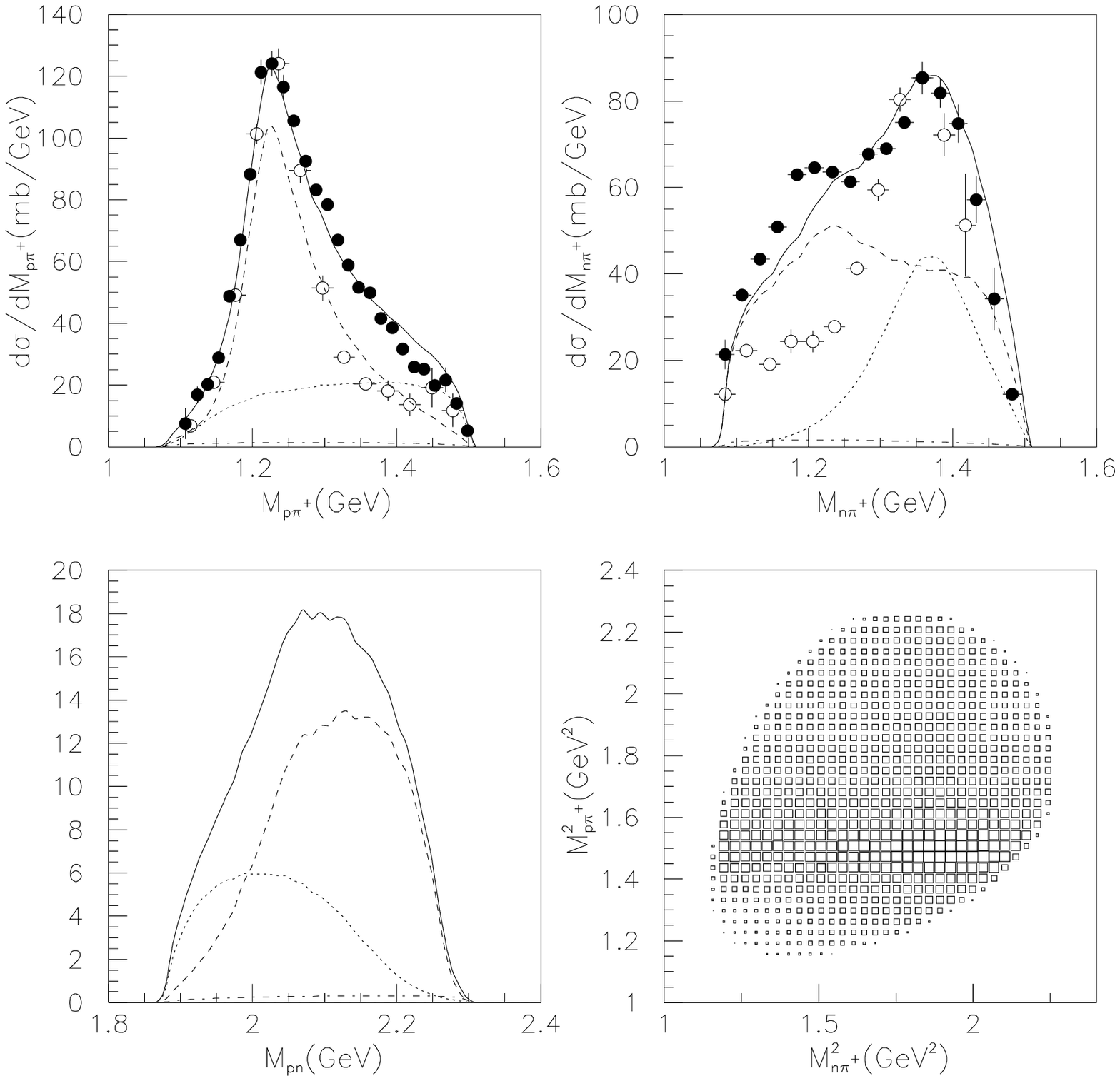}%
\caption{}%
\label{pnpi}%
\end{center}
\end{figure}

The invariant mass spectra and Dalitz plot for the final particles
of $pp \to pn\pi^+$ reaction at T$_\text P=1.3$ GeV are also
calculated as shown in Fig.\ref{pnpi}. While the $\Delta^{++}(1232)$
peak dominates the $p\pi^+$ invariant mass spectrum, a peak due to
$N^*(1440)$ is clearly visible in the $n\pi^+$ invariant mass
spectrum. As a rough comparison with data, preliminary results with
two different triggers and without proper acceptance correction from
CELSIUS-WASA Collaboration~\cite{celsiusroper,nstar2007} are also
shown in the $p\pi^+$ and $n\pi^+$ invariant mass spectra.
Individual contributions to the invariant mass spectra from
$\Delta(1232)$, $N^*(1440)$ and nucleon pole are also given in
Fig.~\ref{pnpi} by dashed, dotted, and dot-dashed curves,
respectively. From the $n\pi^+$ invariant mass spectrum, one can see
that the reflection of $\Delta^{++}(1232)$ resonance gives a large
but flat contribution, the isospin suppressed $\Delta^{+}(1232)$
contribution gives a weak peak around 1.23 GeV, and the Roper
$N^*(1440)$ resonance is absolutely needed to reproduce the peak
around 1.36 GeV. In the calculation for T$_\text P=1.3$ GeV, the
$n$-$p$ FSI factor is not included since the data for both $p\pi^+$
and $n\pi^+$ mass spectra have no evidence for the FSI effect which
should cause further enhancement for both mass spectra at high
energy end. The phenomena may suggest that at beam energy about 1.3
GeV the contribution from $p$-$n$ higher partial waves has already
exceeded the $^3S_1$ partial wave to be the dominant contribution
and the S-wave FSI becomes unimportant. The forthcoming acceptance
corrected results from the CELSIUS-WASA Collaboration and future
experiments at COSY with the newly installed WASA-at-COSY
detector~\cite{adam} or at HIRFL-CSR with the scheduled $4\pi$
hadron detector~\cite{menu2004} will be very helpful for
constraining the ingredients in our model calculations.

In our calculation, the only free parameter we adjust is the
$\Lambda^\Delta_\pi$ for the $\Delta\pi N$ vertex form factor.
Previous calculations used $\Lambda^{\Delta}_{\pi} = 0.65$
GeV~\cite{kundu} and $0.63$ GeV~\cite{dmitriev}. Whereas they
included only the $\Delta$ resonance to reproduce the total cross
sections for the $pp\to pn\pi^+$ reaction, we need to reduce it to
0.59 GeV to allow contributions from the off-shell nucleon pole and
the Roper resonance for the best description of the total cross
section.

The experimental study~\cite{celsiusroper,nstar2007} and our
theoretical study here suggest that the $pp \to pn  \pi^+$ reaction
provides a very good place for studying the $N^*$ resonances. Since
all the contributions from $\Delta^{*+}$ resonances are well
constrained by the corresponding $\Delta^{*++}$ contributions by an
isospin scaling factor of 1/9 and hence suppressed, the $n\pi^+$
invariant mass spectrum for this reaction provides a rather good
isospin filter for $N^*$ resonant peaks. Therefore the study of this
reaction should be extended to higher energies at COSY and HIRFL-CSR
with $4\pi$ detectors to provide various differential cross sections
and the Dalitz plot. It will definitely help us to look for those
``missing" $\Delta^{*++}$ resonances with large coupling to
$\rho^+p$ and $N^*$ resonances with large coupling to $N\sigma$ or
$N\omega$ at higher energies. The information on various couplings
of observed $\Delta^*$ and $N^*$ resonances may also help us gain
some insight on the nature of these resonances~\cite{xie}.

\bigskip
\noindent

{\bf Acknowledgements:} We would like to thank H. Clement, B.C. Liu
and J.J. Wu for useful discussions. This work is partly supported by
the National Natural Science Foundation of China under grants Nos.
10435080, 10521003, 10875133, 10635080, and by the Chinese Academy
of Sciences under projects Nos. KJCX2-SW-N18, KJCX3-SYW-N2,
CXTD-J2005-1.

\newpage
\appendix
\section*{Appendix} In this appendix we present explicitly the
expression of each amplitude in our calculation of Eq.(\ref{ampl})
as the following:
\begin{eqnarray}
{\cal M}(\Delta(1232),\pi^+)& & = \sqrt{2} \frac{f_{\pi N
N}}{m_{\pi}} g^2_{\Delta N \pi}F^{N N}_{\pi}(k^2_{\pi})F^{\Delta
N}_{\pi}(k^2_{\pi}) \bar{u}_3(p_3,s_3) p^{\nu}_{\pi}
G^{\Delta(1232)}_{\nu \mu}(q) k^{\mu}_{\pi}
u_1(p_1,s_1)   \nonumber\\
&& G_{\pi}(k_{\pi}) \bar{u}_n(p_n,s_n)  \gamma_5 \not\! k_{\pi}
u_2(p_2,s_2) - (\text {exchange term with } p_1 \leftrightarrow
p_2), \label{Al}\\
{\cal M}(\Delta(1232),\pi^0)& &=
\frac{\sqrt{2}}{3} \frac{f_{\pi N N}}{m_{\pi}} g^2_{\Delta N
\pi}F^{N N}_{\pi}(k^2_{\pi})F^{\Delta N}_{\pi}(k^2_{\pi})
\bar{u}_n(p_n,s_n) p^{\nu}_{\pi} G^{\Delta(1232)}_{\nu \mu}(q)
k^{\mu}_{\pi}
u_1(p_1,s_1)   \nonumber\\
&& G_{\pi}(k_{\pi}) \bar{u}_3(p_3,s_3)  \gamma_5 \not\! k_{\pi}
u_2(p_2,s_2) - (\text {exchange term with } p_1 \leftrightarrow
p_2),\label{Al}\\
{\cal M}(N^{*}(1440),\pi^0)& &= \sqrt{2} \frac{f_{\pi N N}}{m_{\pi}}
g^2_{N^* N \pi}F^{N N}_{\pi}(k^2_{\pi})F^{N^* N}_{\pi}(k^2_{\pi})
\bar{u}_n(p_n,s_n) \gamma_5 \not \! p_{\pi} G^{N^*(1440)}(q)
\gamma_5 \not\! k_{\pi}
u_1(p_1,s_1)   \nonumber\\
&& G_{\pi}(k_{\pi}) \bar{u}_3(p_3,s_3)  \gamma_5 \not\! k_{\pi}
u_2(p_2,s_2) - (\text {exchange term with } p_1 \leftrightarrow
p_2),\\
{\cal M}(N^{*}(1440),\sigma)& &= \sqrt{2} g_{\sigma N N} g_{N^* N
\sigma} g_{N^* N \pi}F^{N N}_{\sigma}(k^2_{\sigma})F^{N^*
N}_{\sigma}(k^2_{\sigma}) \bar{u}_n(p_n,s_n) \gamma_5 \not \!
p_{\pi} G^{N^*(1440)}(q)
u_1(p_1,s_1)   \nonumber\\
&& G_{\sigma}(k_{\sigma}) \bar{u}_3(p_3,s_3) u_2(p_2,s_2) - (\text
{exchange term with } p_1 \leftrightarrow p_2),\\
{\cal M}(p(938),\pi^0)& &= \sqrt{2} (\frac{f_{\pi N N}}{m_{\pi}})^3
(F^{N N}_{\pi}(k^2_{\pi}))^2F(q) \bar{u}_n(p_n,s_n) \gamma_5 \not \!
p_{\pi} G^{N}(q) \gamma_5 \not\! k_{\pi}
u_1(p_1,s_1)   \nonumber\\
&& G_{\pi}(k_{\pi}) \bar{u}_3(p_3,s_3)  \gamma_5 \not\! k_{\pi}
u_2(p_2,s_2) - (\text {exchange term with } p_1 \leftrightarrow
p_2),\\
{\cal M}(p(938),\sigma)& &= \sqrt{2} \frac{f_{\pi N N}}{m_{\pi}}
g^2_{\sigma N N} (F^{N N}_{\sigma}(k^2_{\sigma}))^2 F(q)
\bar{u}_n(p_n,s_n) \gamma_5 \not \! p_{\pi} G^{N}(q)
u_1(p_1,s_1)   \nonumber\\
&& G_{\sigma}(k_{\sigma}) \bar{u}_3(p_3,s_3) u_2(p_2,s_2) - (\text
{exchange term with } p_1 \leftrightarrow p_2)
\end{eqnarray}
with $u_n(p_n,s_n)$, $u_3(p_3,s_3)$, $u_1(p_1,s_1)$, $u_2(p_2,s_2)$
the spin wave functions of the outgoing neutron, proton in the final
state and two initial protons, respectively. $p_{\pi}$ and
$k_{\pi(\sigma)}$ stand for the 4-momenta of the outgoing
$\pi$-meson and the exchanged $\pi(\sigma)$-meson, respectively.
$p_{1}$ and $p_{2}$ represent the 4-momenta of the two initial
protons.

The amplitudes corresponding to the ``pre-emission" graphs are not
given here, but are included in actual calculations. With these
amplitudes, the $\sum_{s_i} \sum_{s_f} |{\cal A}|^2$ in
Eq.(\ref{eqcs}) is computed with the standard MATHEMATICA package.
\newpage

Table 1: Relevant paramters of $\Delta(1232)$ and $N^*(1440)$.

Figure 1: Feynman diagrams for $pp \to pn  \pi^+$ reaction. Note
that there are also the pre-emission counter parts of these diagrams
where pion is emitted before collision. These graphs are not shown
here, but included in the calculations.

Figure 2: Total cross section vs T$_\text P$ for the $pp \to pn
\pi^+$ reaction from our calculation compared with
data~\cite{data1,data2}. The solid and dashed lines represent the
results with and without $^3S_1$ $np$ FSI, respectively.

Figure 3: Contributions of various components and their simple
incoherent sum as a function of T$_\text P$ for the $pp \to pn
\pi^+$ reaction compared with data~\cite{data1,data2}, without
including $np$ FSI. The dotted, short-dotted, dot-dashed,
dot-dot-dashed, dashed, and short-dashed curves represent
contributions from $\Delta(1232)(\pi^+$ and $\pi^0$ exchange),
$N^*(1440)(\sigma$ and $\pi^0$ exchange) and nucleon pole ($\sigma$
and $\pi^0$ exchange), respectively; the solid curve represents
their simple incoherent sum.

Figure 4: Invariant mass spectra and Dalitz plot for the $pp \to pn
\pi^+$ reaction at T$_\text P=1.3$ GeV, compared with the
preliminary data with two different triggers and without proper
acceptance correction (open circles from Ref.\cite{celsiusroper} and
solid circles from Ref.\cite{nstar2007} with arbitrary
normalization). The dashed, dotted and dot-dashed lines stand for
individual contributions from $\Delta(1232)$, $N^*(1440)$ and
nucleon pole, respectively, while the solid line represents their
simple incoherent sum. In the Dalitz plot, the size of the squares
indicates the magnitude of the number of events.

\newpage

\begin{table}[htbp]
\caption{ \label{table}}
\begin{center}
\begin{tabular}{ccccc}
\hline \hline Resonances  &Width(MeV) & Decay Channel  & Branching
ratios(\%)
& $g^2/4 \pi$  \\
\hline $\Delta^*(1232)$ & 118 & $\pi N$ & 100 &19.54\\
$N^*(1440)$ & 300 & $\pi N$ &65 &0.51 \\
& & $\sigma N$ &7.5 &2.84\\
\hline \hline
\end{tabular}
\end{center}
\end{table}

\newpage

\newpage

\newpage

\newpage

\end{document}